# Implementation of Multipath and Multiple Description Coding in OLSR

Jiazi Yi, Eddy Cizeron, Salima Hamma, Benoît Parrein and Pascal Lesage*
Université de Nantes, Nantes Atlantique Universités
IRCCyN, CNRS UMR 6597, Polytech'Nantes, rue Christian Pauc - BP50609
44306 Nantes cedex 3 – France
{jiazi.yi, eddy.cizeron, salima.hamma, benoit.parrein}@univ-nantes.fr
* Société Keosys, 1, impasse Augustin Fresnel, 44815 Saint Herblain, France {ple@keosys.com}

*Abstract* —In this paper we discussed the application and the implementation of multipath routing and multiple description coding (MDC) extension of OLSR, called MP-OLSR. It is based on the link state algorithm and employs periodic exchange of messages to maintain topology information of the networks. In the mean time, it updates the routing table in an on-demand scheme and forwards the packets in multiple paths which have been determined at the source. If a link failure is detected, the algorithm recovers the route automatically. Concerning the instability of the wireless networks, the multiple description coding is used to improve reliability of the network transmission, and several methods are proposed to allocate the redundancy in different paths. The simulation in NS2 shows that the new protocol can effectively improve the performance of the networks. The implementation of MP-OLSR is also proposed in the end.

## I. INTRODUCTION

In recent years, more and more multipath routing protocols are proposed. These protocols consist of finding multiple routes between a source and destination node. The multipath routing could offer several benefits: load balancing, fault-tolerance, higher aggregate bandwidth, lower end-to-end delay, effectively alleviate congestion and bottlenecks [1] and security.

In the literature, multipath routing protocols are often used for backup routes. Otherwise, if the goal is the repartition of information, the implementation is generally based on pure source routing. Several multipath protocols have been introduced in the work of [6] and [7]. And a source routing multipath OLSR is presented in [8] by using the shortest path algorithm. However, the suppression of nodes in multiple calls of Dijkstra algorithm could not work for sparse networks. Furthermore, strict node-disjoint multiple paths are not suitable for partition or fusion of group of nodes that can imply temporary a single link for connection.

In this study, we discuses a new multipath routing protocol called MP-OLSR based on OLSR[2][3] to provide fault-tolerance, higher aggregate bandwidth and load balancing. Furthermore, the multiple description coding (MDC) provides an additional benefit to the multipath routing. By adding some overheads to each packet, which is calculated as a linear function of the original packets, the resulting packets are geometrically projected into smaller blocks and distributed into available paths. With MDC, given the failure probabilities of the paths, it is possible to find the optimal way to fragment and then distribute the blocks to the paths so that the probability of reconstructing the original information at the destination is maximized.

The remainder of the paper is organized as follows. The specification of the protocol is presented in Section II in detail. In Section III we introduce the application of Multiple Description Coding in the MP-OLSR. The simulation model and performance results are demonstrated in Section IV, and a real implementation is studied in Section V. At last, we conclude the paper in Section VI.

## II. MP-OLSR SPECIFICATION

The MP-OLSR can be regarded as a hybrid multipath routing protocol. It sends out *HELLO* messages and *TC* messages periodically to be aware of the network topology, just like OLSR. However, MP-OLSR does not always keep a routing table. It only computes the routes when there are data packets need to be sent out.

The core functioning of MP-OLSR has two main parts: *topology sensing* and *routes computation*. The *topology sensing* is to make the nodes get to the topology information of the network, which includes *link sensing, neighbor detection* and *topology discovery*. This part gets benefit from *MPRs* as well as OLSR. The *routes computation* uses the *Multipath Dijkstra Algorithm* to populate the multipath based on the information get from the

This work is supported by the French program RNRT (Réseau National de Recherche en Télécommunications) under the project SEREADMO (http://www.sereadmo.org).
This is a proceeding of 4[th] OLSR Introp/Workshop, Ottawa, Canada.

*topology sensing*. The source route (the hops from the source to the destination) will be saved in the header of the data packets. The intermediate nodes just read the packet header and forward the packet to the next hop. Furthermore, to overcome some drawbacks of the source routing, the route recovery is introduced.

*A. Topology Sensing*

The topology sensing is to make the nodes get to know the topology information of the network, which includes link sensing, neighbor detection and topology discovery. This part gets benefit from MPRs as well as OLSR to minimize the flooding of broadcast packets in the network by reducing duplicate retransmissions in the same region.

An adaptation of OLSR for the multipath routing is that in the TC message, the protocol not only include the links between local node and MPR, but the links to all the neighbors so that each node can have better information about the networks topology to construct disjoint multipath. In the simulation, it works well at low data rate. But in the scenario of high data rate, it will cause more congestion because this method will increase the size of TC message. However, with the message combining mechanism in OLSRv2, this problem can be eased.

*B. Routes Computation*

Contrary to classical OLSR, routes are not renewed each time a node receives a new routing message, but in an on-demand scheme, in order to avoid the loud computation of several routes for every possible destination. When a given source must send packets, the route computation procedure uses the algorithm shown in Figure 1.

The general principle of this algorithm is at step *i* to look for the shortest path $P_i$ to the destination *d*. Then the edges in $P_i$ or pointing to $P_i$ have their cost increased in order to prevent the next steps to use similar path. $f_p$ is used to increase costs of arcs belonging to the previously path $P_i$ (or which opposite arcs belong to it). This encourages future paths to use different arcs but not different vertices. $f_e$ is used to increase costs of the arcs who lead to vertices of the previous path $P_i$. We can choose different $f_p$ and $f_e$ to get link-disjoint path or node-disjoint routes as necessary.

In Figure 1, *Dijkstra(G,n)* is the standard Dijkstra's algorithm which provides the source tree of shortest paths from vertex *n* in graph *G*; *GetPath(ST,d)* is the function that extracts the shortest-path to *n* from the source tree *ST*; *Reverse(e)* gives the opposite edge of *e* ; *Head(e)* provides the vertex edge *e* points to.

Another possible solution is instead of increasing the cost of the arc, we delete the related node from the node set so that we can get totally node-disjoint routes, which is the ideal case [8]. However, in the real scenario, especially the cases that the nodes are sparse, the delete of some "key" nodes might prevent from finding new route.

Figure 2 gives the simulation results of the algorithm in the scenario of 300 nodes. As shown in the figure, by using the Multipath Dijkstra Algorithm, we can get three node-disjoint routes. When the number of routes required comes to ten, we get ten optimal routes (some nodes might be shared when node-disjoint routes are unavailable) according to the cost functions. In this article, only three or four routes are used.

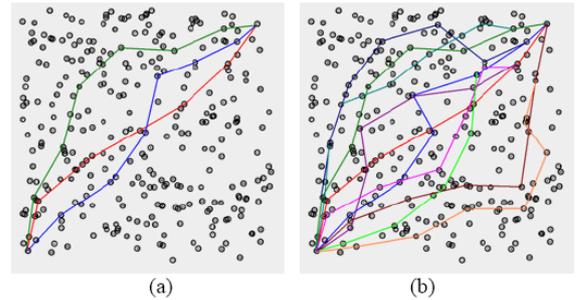

**Figure 1 The Multipath Dijkstra Algorithm**

**Figure 2 Multiple Dijkstra Algorithm with $f_p(c) = f_e(c) = 2c$ (a) Three routes (b) Ten routes**

*C. Route Recovery*

In the classical OLSR, the hop-by-hop routing is used, which means when a packet reaches an intermediate node, the protocol will check the routing table of the local node and then forward the packet to the next hop.

In contrast, in MP-OLSR, we use the semi-source routing approach. It will help the source node keep good control of the packets which will be forwarded in the multipath. However, in the mean time, the pure source routing might cause two problems: Firstly, the information in the source node might be not new enough because it needs time to flood the topology control messages to the whole network. It means when computing the routes, the source node might use the links that does not exist anymore. Secondly, even when the information in the source node is updated, the topology might change during the forwarding of the packet. Both of them will cause the failure of the packets forwarding.

To solve these problems, the *route recovery* is used: before a medium node trying to forward the packet, the node first check if the next hop in the source route of the packet is one of its neighbors. If yes, the packet is forwarded as it should be. If no, then it's possible that the "next hop" has moved out of the transmission range of the node. Then it is necessary to recompute the route and forward the packet through the new route.

### III. THE MULTIPLE DESCRIPTION CODING

By adding redundancy to information streams and splitting them up into several sub-streams, we can improve the integrity of data, especially by sending these sub-streams along different paths from the source to the destination. This kind of transformation is called Multiple Description Coding (see [9] for a detailed review).

Given a piece of information $I$, a multiple description coding method generates $N$ independently communicable packets ($P_1, P_2,..., P_N$). Each description $P_i$ is generally much smaller than the original information. In the MDC scheme, the more descriptions are received, the closer to $I$ is the reconstructed information $\hat{I}$. In our case, we just assume that it exists an integer $M$ ($0<M\leq N$) such that every subset of descriptions containing at least $M$ different descriptions is sufficient to rebuild entirely $I$. The scheme can be easily extended to gracefully reconstruction suitable for image and video services [14]. Thus, the higher is $M$, the lower is the redundancy. In particular, $M = 1$ (respectively $M = N$) corresponds to the case where $(P_i)_{i \in [1,N]}$ are copies of $I$ (respectively where $(P_i)_{i \in [1,N]}$ are different pieces of $I$).

#### A. Mojette Transform

In MP-OSLR, the Mojette transform [10] is used to produce different projections of the original information, each one being sent along a specific path. This discrete form of the Radon transform only requires the addition operation and is exactly invertible. The linear integration of the discrete 2D function $f(k,l)$ is obtained via the Mojette transform over a set of $I$ pre-defined angles, $I = \tan^{-1}(\frac{q_i}{p_i})$. The pairs of integers defining the angles, $(p_i, q_i) = q$, and since linear integration is directionally independent, $q_i$ is restricted to $\mathbb{Z}$ to ensure $q_i \in [0, p]$. Assuring a Dirac pixel model the linear integrations become sums over the pixels centered on the lines $b = q_i k - p_i l$. The Mojette projection operator is defined as

$$M\{f(k,l)\} = Proj(p_i, q_i, b) = \sum_{k=0}^{P-1}\sum_{l=0}^{Q-1} f(k,l)d(b + p_i l - q_i k)$$

where $d(h)$ is the Krönecker function, i.e., $d(h) = 1$ if $h = 0$, otherwise $d(h) = 0$. An example of projections is given in Figure 3. The inverse transform can use Classical Mojette Inversion (CMI) method. For a detailed review, please refer to [11]. Both direct and inverse Mojette operators have linear complexity in number of projections $N$ and number of information elements $I$, i.e. in O($IN$).

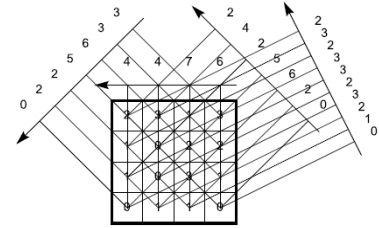

**Figure 3 Four projections of a 4*4 image f(k,l), Proj(-1,1,b), Proj(0,1,b), Proj(1,1,b) and Proj(2,1,b)**

#### B. Geometrical Buffer

However, applying MDC coding to original packets may significantly increase the total number of packets that are transmitted in the network, which may results in new congestion. A possible solution consists in setting up a sending buffer and then performing the MDC on its content (that corresponds to a group of original packets). Furthermore, it could be suitable that MDC is in s systematic construction, which means that first $M$ packets are exact copy of input packets. Incoming packets are thus geometrically buffered and redundant projections are computed. Systematic construction means that when no packet loss occurs, the complexity is null at the decoding side.

#### C. Redundancy Allocation

The redundancy of MDC must be adapted to the transmission environment. It is preferred to transmit packets with more redundancy along the more stable paths to guaranty the final decoding. The metrics can be measured in different layers of the OSI model.

In the physical layer, the link quality based on Bit Error Rate (BER) can be measured and propagated in the network to help decide the cost of the links for the multipath Dijkstra Algorithm. For simulation purposes, it assumes a precise model of propagation that can be given by ray tracing model [16].

In the networks layer, we can monitor the length of the FIFO buffer $B$ at each node and the redundancy is related to possible congestion status. In [15], a simple scheme can improve efficiently the packet delivery ratio but for one path and one hop i.e. the access point. Confident by this result, we propose a heuristic, which is very closed but based on the maximum length of the buffers monitored for each path. Given the max redundancy $K_{max}$, the redundancy to be applied to $M$ different paths is

$$K_i = \frac{1 - \frac{\max(B_i)}{B_{max}}}{\sum_{i=1}^{m}(1 - \frac{\max(B_i)}{B_{max}})} K_{max}$$

where max($B_i$) is the longest buffer length in path $i$ and $B_{max}$ is the buffer size.

Another redundancy allocation scheme is introduced in[14] at the application layer dedicated to image and video transmission, which is to maximize the expected quality $E[Q]$ at the decoding stage:

$$E[Q] = \sum_{j=0}^{L} \Delta Q_j Prob[X \geq r_j]$$

where $\Delta Q$ is the quality increment, $X$ is the random variable representing the number of received projections and $r$ is the function that maps the sub-streams to the corresponding number of projections needed for reconstruction. The allocation redundancy results from the maximization of the expected quality. Metric closed to the human perception are used in [14].

## IV. SIMULATION AND PERFORMANCE ANALYSIS

### A. Environment and Assumption

The proposed algorithm is simulated on NS2. The channel capacity of mobile hosts was set to 11Mbps. A two-ray ground reflection model, which considers both the direct and a ground reflection path, was used as radio propagation model. We use the DCF (distributed coordination function) of IEEE 802.11 for wireless LAN as the MAC layer protocol. It has functionality to notify the network layer about link breakage. In the simulations, there are 50 nodes move in a 1000m×1000m square region and 30 CBR sources are included in the transmission.

### B. Simulation results

We compared the performance of OLSR and MP-OLSR in the simulations. Except for the difference of the protocols, two different kinds of strategy to discover the link failures are considered. The first one is the proactive way: when a node does not receive *HELLO* messages continuously from its former neighbors, it considers that the wireless link between them is broken and will delete the corresponding node from the neighbor set. The second strategy is to use link-layer feedback. When a node is trying to forward a packet to the next hop that has been out of its transmission range, its link layer will drop the packet because of *MAC_RET (*REtry Timeout). In the mean time, the link layer will give a feedback to the routing layer and inform the lost of the link. We also compared both of the strategies in the simulation.

The simulations are taken in 4 protocols:
- *The original OLSR*: It is the original single path routing protocol. It discovers the link failure in a proactive way. The implementation of UM-OLSR [12] is used here;
- *OLSR with link layer feedback* [12];
- *SR-MPOLSR: MP-OLSR* with link layer feedback, use source routing only. The incremental functions are: $f_p(c) = f_e(c) = 2c$;
- *RE-MPOLSR:* MP-OLSR with route recovery and link layer feedback. The incremental functions are: $f_p(c) = f_e(c) = 2c$.

Figure 4 shows the delivery ratio of the simulations. As we can see from the figure, the OLSR with feedback has better delivery ratio than the original OLSR. This is because the protocol with feedback could detect a link failure as soon as the first packet is dropped and recomputed the routing table. In the contrast, the original OLSR tends to continue to send the packets through a failed link until it finds out that it is not able to receive a *HELLO* message from the original neighbor. Both with the link layer feedback, the SR-MPOLSR's delivery ratio is about ten percent lower than the OLSR. This is because of the drawbacks of source routing that have been mentioned in the previous section. However, the RE-MPOLSR gives better delivery ratio thanks to the route recovery.

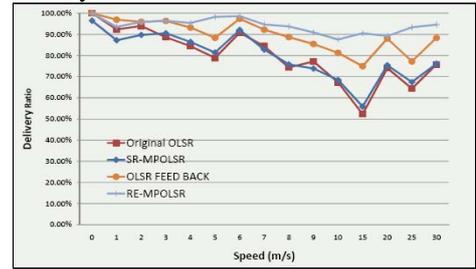

**Figure 4 Delivery Ratio**

Figure 5 shows the average end-to-end delay. It includes the queue delay in every node and the propagation delay from the source to the destination. The multipath routing could effectively reduce the queue delay because the traffic is distributed in different routes and gives the minimum delay.

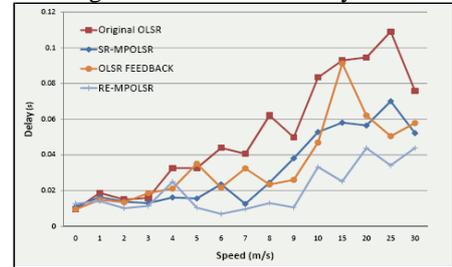

**Figure 5 End to end delay**

The simulation of MP-OLSR with MDC is also taken. The gain of delivery ratio is very little (one or two percent) compared with the RE-MPOLSR in the scenario of 50 nodes. This is because given the low density of the nodes, sometimes the multiple paths will share a lot of links or even there is just no route at all. Then it is hard to make benefits from the MDC. In fact, the MDC tends to be more suitable for large and dense networks.

Figure 6 gives the delivery ratio in the scenario of 100 nodes. This figure compares 3 protocols: RE-MPOLSR, OLSR Feedback and MDC-OLSR in which the number of descriptions is 4 (i.e. number of routes is 4) and the number of useful descriptions in the reconstruction of initial information is 2. So with higher density of the nodes, MP-OLSR with MDC will have more reliable data transmission by offering better delivery ratio. It is about 10% for speed range between 6m/s and 10m/s.

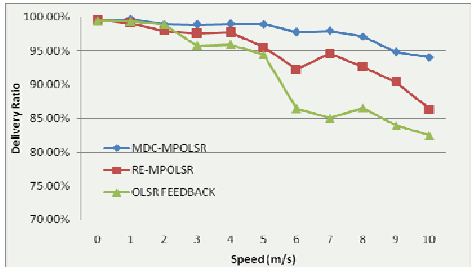

**Figure 6 Delivery ratio in the scenario of 100 nodes**

## V. IMPLEMENTATION OF MP-OLSR

The real implementation is also taken into consideration on the Linux platform based on olsrd [13]. The OLSR protocol acts as an application upon the Linux kernel and always provides the next hop information to the destination for the IP routing protocol. However, in MP-OLSR, it is possible to have different next hop for the same destination. So we modified the IP routing: whenever there is a route request, it will trigger the MP-OLSR to provide the information of the next hop, and this information will be injected to the IP routing table. Furthermore, the Mojette transform and the associated security protocol SERANO[17] is also located in the protocol stack, as shown in

Figure 7. The anonymous forwarding for multiple hops (particularly the source and the destination), the integrity control and the cypherment procedures can be provided by SERANO stack. The choice of entire security solution needs to switch off route recovery.

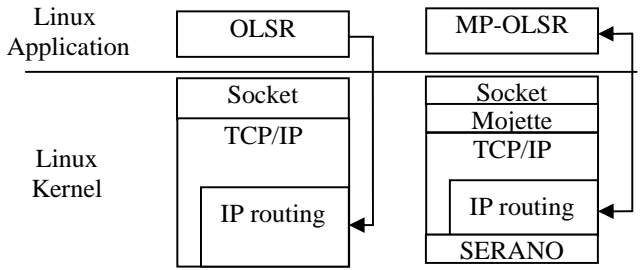

**Figure 7 OLSR and MP-OLSR in Linux**

## VI. CONCLUSIONS AND FUTURE WORKS

In this paper, we discussed a multipath extension to OLSR. Our Multipath Dijkstra Algorithm is used to discover the disjoint routes and the source routing with route recovery is performed to forward the packets. In addition, the MDC is applied to enhance the reliability of the transmission. Security functions can be implemented via dedicated SERANO stack.

The future research includes refining the incremental functions of $f_p$ and $f_e$ to make them adaptive to the specific network and optimize the redundancy allocation for MDC in the data transmission. More largely, more precise simulations of physical layer within NS2 are engaged.